\documentclass{icrc2009}

\usepackage{graphicx}   
\usepackage{caption}    
\usepackage[font=footnotesize]{subfig} 
\usepackage{fixltx2e}
\usepackage{url}

\def\apj{The Astrophysical Journal}%
\def\aap{Astronomy \& Astrophysics}%

\newcommand{\shorttitle}[1]%
{\markboth{Proceedings of the 31\MakeLowercase{$^{st}$} ICRC, {\L}\'{o}d\'{z} 2009}{#1} }
\newcommand{\etal}{\MakeLowercase{\textit{et al. }}} 
\newcommand{\lsi}    {LS~I~+61~303}
\newcommand{\lsif}   {LS~I~+61~303}
\newcommand{\xmm}    {{\it XMM-Newton}}
\newcommand{\swift}  {{\it Swift}}
\newcommand{\lomb}    {Lomb-Scargle}

\begin{document}
\title{The MAGIC highlights of the gamma ray binary LS I +61 303}

\author{\IEEEauthorblockN{T.~Jogler\IEEEauthorrefmark{1},
                             N.~Puchades\IEEEauthorrefmark{2},
                          O.~Blanch Bigas\IEEEauthorrefmark{2},
                           V.~Bosch-Ramon\IEEEauthorrefmark{3},
                           J.~Cortina\IEEEauthorrefmark{2},
                           J.~Mold\'on\IEEEauthorrefmark{3},
                           J.~M.~Paredes\IEEEauthorrefmark{3},\\
                            M.~A.~Perez-Torres\IEEEauthorrefmark{4},
                            M.~Rib\'o\IEEEauthorrefmark{3},
                            J.~Rico\IEEEauthorrefmark{5}\IEEEauthorrefmark{2},
                            D.~F.~Torres\IEEEauthorrefmark{5}\IEEEauthorrefmark{6},
                            and
                           V.~Zabalza\IEEEauthorrefmark{3}}
                           on behalf the MAGIC collaboration
                            \\
\IEEEauthorblockA{\IEEEauthorrefmark{1}Max-Planck-Institut f\"ur
Physik, D-80805 M\"unchen, Germany}
\IEEEauthorblockA{\IEEEauthorrefmark{2}IFAE, Edifici Cn., Campus
UAB, E-08193 Bellaterra, Spain}
\IEEEauthorblockA{\IEEEauthorrefmark{3}Universitat de Barcelona
(ICC/IEEC), E-08028 Barcelona, Spain}
\IEEEauthorblockA{\IEEEauthorrefmark{4}Inst. de Astrof\'{\i}sica
de Andalucia (CSIC), E-18080 Granada, Spain}
\IEEEauthorblockA{\IEEEauthorrefmark{5} ICREA, E-08010 Barcelona,
Spain} \IEEEauthorblockA{\IEEEauthorrefmark{6} Institut de
Cienci\`es de l'Espai (IEEC-CSIC), E-08193 Bellaterra, Spain} }

\shorttitle{T. Jogler \etal MAGIC observations of \lsi} \maketitle

\begin{abstract}
The discovery of emission of TeV gamma rays from X-ray binaries
has triggered an intense effort to better understand the particle
acceleration, absorption, and emission mechanisms in compact
binary systems. Here we present the pioneering effort of the MAGIC
collaboration to understand the very high energy emission of the
prototype system \lsi. We report on the variable nature of the
emission from \lsi\ and show that this emission is indeed
periodic. The system shows regular outburst at TeV energies in
phase $\phi=0.6-0.7$ and detect no signal at periastron ($\phi\sim
0.275$). Furthermore we find no indication of spectral variation
along the orbit of the compact object and the spectral energy
distribution is compatible with a simple power law with index
$\Gamma=2.6\pm0.2_{stat}\pm0.2_{sys}$. To answer some of the open
questions concerning the emission process of the TeV radiation we
conducted a multiwavelength campaign with the MAGIC telescope,
\xmm, and \swift\ in September 2007. We detect a simultaneous
outburst at X-ray and TeV energies, with the peak at phase 0.62
and a similar shape at both wavelengths. A linear fit to the
strictly simultaneous X-ray/TeV flux pairs provides
$r=0.81_{-0.21}^{+0.06}$. Here we present the observations and
discuss the implications of the obtained results to the emission
processes in the system.

  \end{abstract}


\section{Introduction}
\label{introduction}
\lsi\ is a high mass X-ray binary system located at
2.0$\pm$0.2~kpc from us \cite{1991AJ....101.2126F}. The system
contains a rapidly rotating early type B0\,Ve star with a stable
equatorial decretion disk and mass loss, and a compact object with
a mass between 1 and 4~M$_\odot$ orbiting it every $\sim$26.5~d
(see \cite{Casares:2005wn},
\cite{optical_lsi_grundstrom_I2007ApJ...656..437G},
\cite{lsi_orbit_aragona2009AAS...21341008A}, and references
therein). Although \lsi\ has been classified as a microquasar
\cite{radio_lsi_precessing_jet_massi2004A&A...414L...1M}, VLBA
images obtained during a full orbital cycle show an elongated
morphology that rotates as a function of the orbital phase
\cite{2006smqw.confE..52D}. Later VLBA images show repeating
morphologies at the same orbital phases, suggesting that the
milliarcsecond morphology depends only on the orbital phase
\cite{MAGIC_lsi_mw_2008ApJ}. This may be consistent with a model
based on the interaction between the relativistic wind of a young
non-accreting pulsar and the wind/decretion disk of the stellar
companion \cite{Dubus2006_ab}\\
\lsi\ shows periodic non-thermal radio outbursts on average every
$P_{\rm orb}$=26.4960$\pm$0.0028~d, with the peak of the radio
emission shifting between phase 0.45 and 0.95, using
$T_0$=JD~2,443,366.775, in a superorbital period of 1667$\pm$8~d
\cite{radio_lsi_period_best2002ApJ...575..427G}. According to the
most precise orbital parameters periastron takes place at phase
0.275 and the eccentricity of the orbit is $0.537\pm0.034$
\cite{lsi_orbit_aragona2009AAS...21341008A}. \\
\lsi\ has been observed several times in the X-ray domain (see
\cite{2009ApJlsi_xray_smith} and references therein). It generally
displays
 X-ray outbursts, between orbital phase 0.4 and 0.8.\\
At very high energy (VHE) gamma rays \lsi\ has been extensively
studied by MAGIC \cite{Albert:2006vk,2009ApJ_lsi_magic_period} and
VERITAS \cite{2008ApJ...679.1427A}. The lack of a systematic
behavior from cycle to cycle at X-ray energies, and the occurrence
of short-term variability, prevents to establish a X-ray/TeV
correlation from the comparison of non-simultaneous data~\cite{2009arXiv0904.4422V}.\\
Here we report about the highlights of the MAGIC VHE gamma ray
observation and present our strictly simultaneous TeV and X-ray
observations of \lsi.

\section{VHE Gamma Ray observations and data analysis}
\label{observations}
The MAGIC telescope located on the Canary Island of La Palma
($28.75^\circ$N, $17.86^\circ$W, 2225~m a.s.l.). Its essential
parameters are a 17~m diameter segmented mirror of parabolic
shape, an $f/D$ of 1.05 and an hexagonally shaped camera of 576
hemispherical photo multiplier tubes with a field of view of
$3.5^\circ$ diameter. MAGIC can detect gamma rays from 60~GeV to
several TeV. Its energy resolution is $\Delta E=20$\% above
energies of 200~GeV. The current sensitivity is 1.6\% of the Crab
Nebula flux for a $5\sigma$ detection in 50~h of observation time.
The improvement compared to previous sensitivity was achieved by
installing new 2~GHz FADCs \cite{2009_magic_timing_improv}.

The data analysis was carried out using the standard MAGIC
analysis and reconstruction software \cite{2008ApJ...674.1037A}
and is in detail described
in~\cite{Albert:2006vk,2009ApJ_lsi_magic_period,2009ApJ_lsi_correlation}.

The TeV observations were performed in three distinct
observational campaigns (OC hereafter). OC~I, which lead to the
discovery of \lsi\ as a $\gamma$-ray emitter, was performed from
September 2005 to March 2006 and covered 6 orbital periods of
\lsi\ with a total effective observation time of 54~h. OC~II
covered only 4 orbital periods but with a much denser sampling
compared to OC~I and resulted in an effective observation time of
112~h taken in September 2006 to December 2006. Finally OC~III
data were taken from 4th - 21st September 2007 with an effective
observation time of 54~h. The range of zenith angles for all
observations was $[32^{\circ},55^{\circ}]$, although most of the
data had zenith angle below $44^{\circ}$. More details about these
observations can be found
in~\cite{Albert:2006vk,2009ApJ_lsi_magic_period,2009ApJ_lsi_correlation}.

\section{The VHE gamma ray temporal behavior}
\begin{figure}[tbp]
  \centering
  \includegraphics[width=\linewidth]{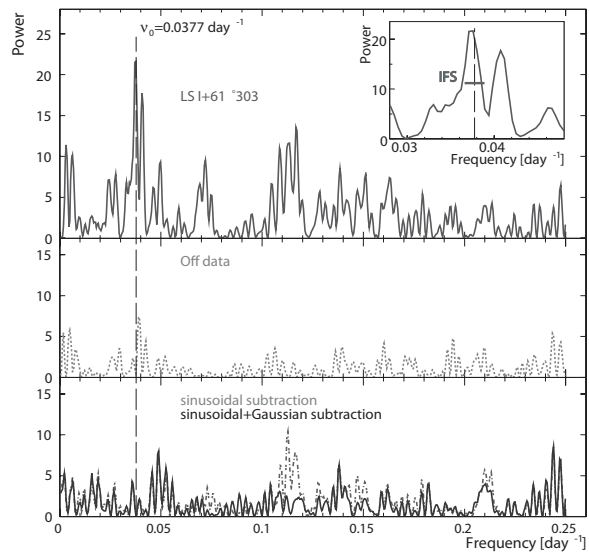}
  \caption{\lomb\ periodogram over the combined
    2005 and 2006 campaigns of \lsi\ data (upper panel) and simultaneous
    background data (middle panel).
    In the lower panel we show the periodograms after subtraction of a
    sinusoidal signal at the orbital
    period (doted) and a sinusoidal plus a Gaussian wave form
    (solid).
    Vertical dashed line corresponds to the orbital frequency.
    Inset: zoom around the highest peak, which corresponds to the
    orbital frequency ($0.0377$d$^{-1}$).
    Its post-trial probability is nearly $10^{-7}$ The IFS is also shown.}
  \label{fig:period}
\end{figure}

The light curve in OC~I and OC~II is derived above $E>400\textrm{
GeV}$ and for OC~III above $E>300\textrm{ GeV}$ due to the
improved sensitivity.\\
The most pronounced feature in the light curve is the high flux
level in the phase range $\phi$~0.6-0.7. In this phase range is
almost every time the
highest flux (during one orbital cycle) measured.\\
Since the real value of the periastron passage is yet debated the
averaged flux value between the phase bin 0.2--0.3 can be used as
an upper limit to the emission at periastron. Thus the flux must
be less than $F(\mathrm{E}>400 \mathrm{ GeV})=2.2 \times 10^{-12}
\mathrm{cm}^{-2}\mathrm{s}^{-1}$ at the $95\%$ confidence level.\\
Additional significant fluxes are measured during phase 0.5-0.6 in
OC~II and OC~I. In OC~II and in OC~III additional high fluxes are
evident in the phase range 0.8-1.0. In OC~II only a single
significant measurement occurs at $\phi=0.84$, while the emission
spreads over more nights in OC~III yielding an averaged
significant ($\sim 5\sigma$) signal in the phase range 0.8-1.0 at
the level of $(5.2\pm1.0)\times10^{-12}$~cm$^{-2}$~s$^{-1}$. This
flux level is compatible with the 2-$\sigma$ upper limit we obtain
for the OC~II~\cite{2009ApJ_lsi_magic_period}. The higher
sensitivity both due to improved hardware and longer observation
times can sufficiently
explain the non detection in OC~II.\\
We performed a test for periodicity applying the \lomb\
\cite{1982ApJ...263..835S} method. The data from OC~I and OC~II
are used and the light curve is binned in intervals of $\Delta
t=15\textrm{ min}$ to assure that each measurement has the same
error and thus can be treated equal. To evaluate the complementary
cumulative probability function we performed a monte carlo
simulation by generating $10^6$ random generated light curves
(from gaussian white noise) with the same sampling as the \lsi\
data. For each of the $10^6$ light curves the \lomb\ periodogram
is calculated and from the maximum peak distribution of all
periodograms the complementary cumulative probability density
function (cCPF) is computed. The periodogram obtained from the
\lsi\ light curve is shown together with the background event
periodogram in Fig.\ref{fig:period}. A highly significant peak is
found at a period of $P=26.8\pm0.2\textrm{ d}$ with a false alarm
probability of $\sim 10^{-7}$ in very good agreement with the
orbital period of the system. Our test proves that the VHE
emission from \lsi\ is periodic modulated with the orbital period.
A very detailed description of our periodicity analysis and the
treatment of possible systematic effects of the method can be
found in~\cite{2009ApJ_lsi_magic_period}.

\section{The VHE spectral behavior}
The flux measured from \lsi\ in individual nights is usually not
high enough to obtain significant spectra. The only exception is
the main emission peak in phase $\phi$~0.6-0.7. We obtained
several spectra for individual phase bins in each OC. All spectral
energy distributions are compatible with simple power laws. Within
the errors all obtained fit parameters are compatible with each
other and with the most significant measured spectrum
$\frac{\mathrm{d}F}{\mathrm{d}E} =
\frac{(1.2\pm0.4_{\mathrm{stat}}\pm0.3_{\mathrm{syst}}) \cdot
10^{-12}} {\mathrm{TeV}\,\mathrm{cm}^2\,\mathrm{s}}
\left(\frac{E}{1\,
\mathrm{TeV}}\right)^{-2.7\pm0.4_{\mathrm{stat}}\pm0.2_{\mathrm{syst}}}
$. So no significant spectral variation could be found. We
investigated the less significant flux measurements by calculating
a hardness ration ($HR$), which we define as the ratio of the
integral flux between 400~GeV and 900~GeV and  above 900~GeV. We
do not find any correlation between the $HR$ and the flux level.

\section{Multiwavelength campaign}
\label{Multiwavelength campaign}
Multiwavelength observations during OC~II (see
\cite{MAGIC_lsi_mw_2008ApJ}) yielded no correlation between the
radio and the TeV emission of \lsi. Hints of correlated X-ray/TeV
emission have been found  based on non-simultaneous data taken
more than six hours \cite{2009ApJ_lsi_magic_period} and one day
apart in these observations. Here we report only on the strictly
simultaneous data taken in 2007.

\subsection{X Ray observations} \label{obs_xray}
We observed \lsi\ with \xmm\ during seven runs from 2007 September
4 to 11, amounting to a total observation time of 104.3~ks. The
data were processed using the version 8.0.0 of the \xmm\ Science
Analysis Software (SAS). Known hot or flickering pixels were
removed using the standard SAS tasks. Further cleaning to remove
from the dataset periods of high background reduced the net total
good exposure times to 67.0 and
92.6~ks for the pn and MOS detectors, respectively.\\
Source spectra were extracted from a $\sim70"$ radius circle
centered on the source (PSF of $15"$) while background spectra
were taken from a number of source-free circles with $\sim 150"$
radius. The extracted spectra were analyzed with XSpec v12.3.1
\cite{XSPEC1996}. An absorbed power-law function
 yielded satisfactory fits for all
observations. Unabsorbed fluxes in the
0.3--10~keV range were computed from the spectral fits.\\
Additional observations of 2--5~ks each (total 28.5~ks) were
obtained with the \swift/XRT from 2007 September 11 to 22. The
total observation time was 28.5~ks. The \swift\ data were
processed using the FTOOLS task {\tt xrtpipeline}. The spectral
analysis procedures were the same as those used for the \xmm\
data, but fixing the hydrogen column density to
$0.5\times10^{22}$~cm$^{-2}$, a typical value for \lsi\ also found
in the \xmm\ fits.\\
To look for short-term X-ray variability we also extracted
0.3--10~keV background-subtracted lightcurves for each
observation. In addition we computed hardness ratios as the
fraction between the count rates above and below 2~keV. More
information about the X-ray analysis can be found
in~\cite{2009ApJ_lsi_correlation}.

\begin{figure}[tb]
\resizebox{1.0\hsize}{!}{\includegraphics[angle=0]{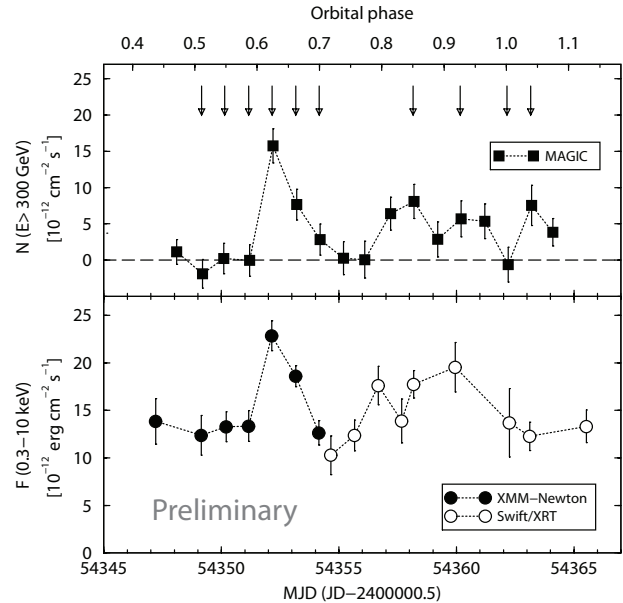}}
\caption{ TeV and X-ray lightcurves of \lsif\ during the
multiwavelength campaign of 2007 September. {\it Top}: Flux above
300~GeV versus the observation time in MJD and the orbital phase.
The horizontal dashed line indicates 0 flux. The vertical arrows
mark the times of simultaneous TeV and X-ray observations. {\it
Bottom}: Unabsorbed flux in the 0.3--10~keV energy range for the
seven \xmm\ observations (filled circles) and the nine \swift\
ones (open circles). Error bars correspond to a 1-$\sigma$
confidence level in all cases. Dotted lines join consecutive data
points to help following the main trends of the lightcurves.
\label{fig:lc}}
\end{figure}

\subsection{X-ray results \& X-ray/TeV Correlation} \label{res_xray}
There is no significant hardness ratio change within each of our
observations. Thus the unabsorbed flux obtained from the spectral
fit is a good estimate of the unabsorbed flux during the
observation. Still moderate ($\Delta F<25\%$) count-rate
variability is present in most observations. We converted this
count-rate variability into flux variability and added  this flux
variability (as an estimate of additional flux uncertainty) in
quadrature to the spectral fits flux errors. This procedure
provides more realistic total flux uncertainties.

We show in Fig.~\ref{fig:lc}-bottom the 0.3--10~keV lightcurve of
\lsi\ together with the VHE lightcurve obtained with MAGIC.
\begin{figure}[tb]
\center
\resizebox{0.9\hsize}{!}{\includegraphics[angle=0]{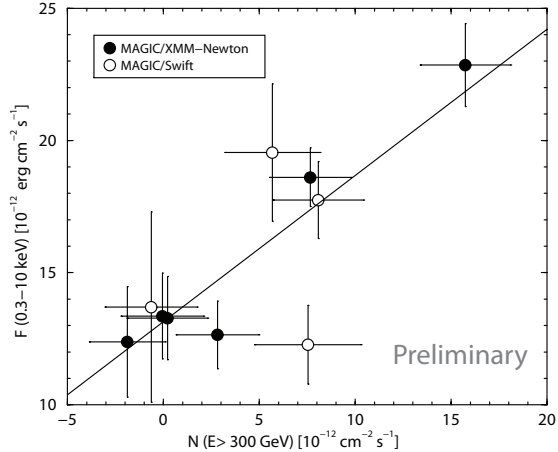}}
\caption{Unabsorbed X-ray fluxes as a function of TeV fluxes for
\lsif\ during the multiwavelength campaign of 2007 September. Only
the 10 simultaneous fluxes, marked with arrows in
Fig.~\ref{fig:lc}, have been considered. Filled circles represent
\xmm\ measurements, while open ones are those from \swift. Error
bars correspond to a 1-$\sigma$ confidence level in all cases. The
solid line represents a linear fit to all data points.
\label{fig:corr}}
\end{figure}
A clear correlation between the X-ray and TeV emissions is
observed with a simultaneous peak at phase 0.62 (see
Fig.~\ref{fig:lc}). We plot in Fig.~\ref{fig:corr} the X-ray
fluxes against the TeV fluxes for strictly simultaneous taken
data, amounting to 10 data points (marked with arrows in
Fig.~\ref{fig:lc}). A linear fit to the six MAGIC/\xmm\ pairs that
trace the outburst yields a correlation coefficient of $r=0.97$. A
linear fit to all ten simultaneous pairs provides a high
correlation coefficient of $r=0.81_{-0.21}^{+0.06}$ (which has a
probability of about $5\times10^{-3}$ to be produced from
independent X-ray and TeV fluxes).\\

Contemporaneous radio data obtained with RATAN, VLBA and H$\alpha$
spectroscopy, are consistent with previous result (details will be
reported elsewhere). Therefore, the X-ray/TeV correlation occurred
when the source was showing a standard behavior in both its
outflow (radio) and decretion disk (H$\alpha$ line).

\section{Conclusion}
\label{discussion}
We find that \lsi\ is a periodic $\gamma$-ray binary with an
orbital period of 26.8$\pm$0.2~days (chance probability $\sim
10^{-7}$), compatible with the optical, radio and X-ray period.
This result implies that the flux modulation is tied to the
orbital period.\\
We produce energy spectra for several phase bins and the spectral
photon index does not show a significant dependence on the orbital
phase.\\
We put constraints to the emission at the periastron passage and
conclude that the system is detected in $\gamma$-rays only in the
phases $0.4-1.0$. Since significant emission is only detected in
an orbital sector off the phases at which the maximum gamma ray
flux should occur under photon-photon absorption, the latter can
hardly be the only source of variability in the emission.

In addition we have discovered an X-ray/TeV correlation in \lsi\
based on simultaneous multiwavelength data obtained with MAGIC,
\xmm, and \swift. The quoted X-ray fluxes are already unabsorbed
 and the TeV spectra show no absorption indication nor is
significant absorption predicted for the explored phase range.
Therefore, the X-ray/TeV correlation we have found for \lsi\
indicates that the emission processes at both wavelengths occur at
the same time and are probably the result of a single physical
mechanism.

Since the VHE flux is about a factor of 2 lower than the X-ray
flux measured the X-ray/TeV correlation favor leptonic models if
the radiation mechanisms are dominated by a single particle
population. In addition, the IC cooling channel is less efficient
than the synchrotron channel to produce the detected X-ray
emission for reasonable values of the magnetic field. This
suggests that the X-rays are the result of synchrotron radiation
of the same VHE electrons that produce TeV emission as a result of
inverse Compton scattering of optical/ultraviolet stellar photons.

\section*{acknowledgments}
We thank N.~Gehrels for his help in arranging the {\it Swift}
observations. We would like to thank the Instituto de Astrofisica
de Canarias for the excellent working conditions at the
Observatorio del Roque de los Muchachos in La Palma. The support
of the German BMBF and MPG, the Italian INFN and Spanish MICINN is
gratefully acknowledged. This work was also supported by ETH
Research Grant TH 34/043, by the Polish MNiSzW Grant N N203
390834, and by the YIP of the Helmholtz Gemeinschaft.



\begin{thebibliography}{10}
\providecommand{\url}[1]{#1} \csname url@samestyle\endcsname
\providecommand{\newblock}{\relax}
\providecommand{\bibinfo}[2]{#2}
\providecommand{\BIBentrySTDinterwordspacing}{\spaceskip=0pt\relax}
\providecommand{\BIBentryALTinterwordstretchfactor}{4}
\providecommand{\BIBentryALTinterwordspacing}{\spaceskip=\fontdimen2\font
plus \BIBentryALTinterwordstretchfactor\fontdimen3\font minus
  \fontdimen4\font\relax}
\providecommand{\BIBforeignlanguage}[2]{{%
\expandafter\ifx\csname l@#1\endcsname\relax
\typeout{** WARNING: IEEEtran.bst: No hyphenation pattern has been}%
\typeout{** loaded for the language `#1'. Using the pattern for}%
\typeout{** the default language instead.}%
\else \language=\csname l@#1\endcsname \fi #2}}
\providecommand{\BIBdecl}{\relax} \BIBdecl

\bibitem{1991AJ....101.2126F}
D.~A. {Frail} and R.~M. {Hjellming}, ``{Distance and total column
density to
  the periodic radio star LSI + 61{$^{\circ}$} 303},'' \emph{\apj}, vol. 101, pp.
  2126--2130, Jun. 1991.

\bibitem{Casares:2005wn}
J.~Casares \emph{et~al.}, ``Orbital parameters of the microquasar
lsi +61
  303,'' \emph{Mon. Not. Roy. Astron. Soc.}, vol. 360, pp. 1091--1104, 2005.

\bibitem{optical_lsi_grundstrom_I2007ApJ...656..437G}
E.~D. {Grundstrom} \emph{et~al.}, ``{Joint H{$\alpha$} and X-Ray
Observations
  of Massive X-Ray Binaries. II. The Be X-Ray Binary and Microquasar LS I +61
  303},'' \emph{\apj}, vol. 656, pp. 437--443, Feb. 2007.

\bibitem{lsi_orbit_aragona2009AAS...21341008A}
C.~{Aragona} \emph{et~al.}, ``{Optical Spectroscopy of Gamma-ray
Binaries},''
  in \emph{American Astronomical Society Meeting Abstracts}, ser. American
  Astronomical Society Meeting Abstracts, vol. 213, Jan. 2009, pp. 410.08--+.

\bibitem{radio_lsi_precessing_jet_massi2004A&A...414L...1M}
M.~{Massi} \emph{et~al.}, ``{Hints for a fast precessing
relativistic radio jet
  in LS I +61{$^\circ$}303},'' \emph{\aap}, vol. 414, pp. L1--L4, Jan. 2004.

\bibitem{2006smqw.confE..52D}
V.~{Dhawan}, A.~{Mioduszewski}, and M.~{Rupen}, ``{LS I +61 303 is
a Be-Pulsar
  binary, not a Microquasar},'' in \emph{Proceedings of the VI Microquasar
  Workshop: Microquasars and Beyond. September 18-22, 2006, Como, Italy.,
  p.52.1}, 2006.

\bibitem{MAGIC_lsi_mw_2008ApJ}
J.~{Albert} \emph{et~al.}, ``{Multiwavelength (Radio, X-Ray, and
  {$\gamma$}-Ray) Observations of the {$\gamma$}-Ray Binary LS I +61 303},''
  \emph{\apj}, vol. 684, pp. 1351--1358, Sep. 2008.

\bibitem{Dubus2006_ab}
G.~{Dubus}, ``{Gamma-ray absorption in massive X-ray binaries},''
\emph{\aap},
  vol. 451, pp. 9--18, May 2006.

\bibitem{radio_lsi_period_best2002ApJ...575..427G}
P.~C. {Gregory}, ``{Bayesian Analysis of Radio Observations of the
Be X-Ray
  Binary LS I +61{$^\circ$}303},'' \emph{\apj}, vol. 575, pp. 427--434, Aug.
  2002.

\bibitem{2009ApJlsi_xray_smith}
A.~{Smith} \emph{et~al.}, ``{Long-Term X-Ray Monitoring of the TeV
Binary LS I
  +61 303 With the Rossi X-Ray Timing Explorer},'' \emph{\apj}, vol. 693, pp.
  1621--1627, Mar. 2009.

\bibitem{Albert:2006vk}
J.~Albert \emph{et~al.}, ``Variable very high energy gamma-ray
emission from
  the microquasar LS I +61 303,'' \emph{Science}, vol. 312, pp. 1771--1773,
  2006.

\bibitem{2009ApJ_lsi_magic_period}
J.~{Albert} \emph{et~al.}, ``{Periodic Very High Energy
{$\gamma$}-Ray Emission
  from LS I +61{$^{\circ}$} 303 Observed with the MAGIC Telescope},''
  \emph{\apj}, vol. 693, pp. 303--310, Mar. 2009.

\bibitem{2008ApJ...679.1427A}
V.~A. {Acciari} \emph{et~al.}, ``{VERITAS Observations of the
{$\gamma$}-Ray
  Binary LS I +61 303},'' \emph{\apj}, vol. 679, pp. 1427--1432, Jun. 2008.

\bibitem{2009arXiv0904.4422V}
{V.~A.~Acciari} \emph{et~al.}, ``{Multiwavelength Observations of
LS I +61 303
  with VERITAS, Swift and RXTE},'' \emph{ArXiv e-prints}, Apr. 2009.

\bibitem{2009_magic_timing_improv}
E.~{Aliu} \emph{et~al.}, ``{Improving the performance of the
single-dish
  Cherenkov telescope MAGIC through the use of signal timing},''
  \emph{Astroparticle Physics}, vol.~30, pp. 293--305, Jan. 2009.

\bibitem{2008ApJ...674.1037A}
J.~{Albert} \emph{et~al.}, ``{VHE {$\gamma$}-Ray Observation of
the Crab Nebula
  and its Pulsar with the MAGIC Telescope},'' \emph{\apj}, vol. 674, pp.
  1037--1055, Feb. 2008.

\bibitem{2009ApJ_lsi_correlation}
------, ``{in preparation}.''

\bibitem{1982ApJ...263..835S}
J.~D. {Scargle}, ``{Studies in astronomical time series analysis.
II -
  Statistical aspects of spectral analysis of unevenly spaced data},''
  \emph{\apj}, vol. 263, pp. 835--853, Dec. 1982.

\bibitem{XSPEC1996}
K.~A. {Arnaud}, ``{XSPEC: The First Ten Years},'' in
\emph{Astronomical Data
  Analysis Software and Systems V}, ser. Astronomical Society of the Pacific
  Conference Series, G.~H. {Jacoby} and J.~{Barnes}, Eds., vol. 101, 1996, pp.
  17--+.

\end{thebibliography}

\end{document}